\documentclass[twocolumn,10pt]{article}

\usepackage[utf8]{inputenc}
\usepackage[T1]{fontenc}
\usepackage{times}
\usepackage{graphicx}
\usepackage{amsmath,amssymb}
\usepackage{booktabs}
\usepackage[margin=2cm]{geometry}
\usepackage{xcolor}
\usepackage[colorlinks=true,linkcolor=blue,citecolor=blue,urlcolor=blue]{hyperref}
\usepackage{listings}
\usepackage{xcolor}

\definecolor{codegreen}{rgb}{0,0.6,0}
\definecolor{codegray}{rgb}{0.5,0.5,0.5}
\definecolor{codepurple}{rgb}{0.58,0,0.82}
\definecolor{backcolour}{rgb}{0.95,0.95,0.92}

\lstdefinestyle{mystyle}{
    commentstyle=\color{codegreen},
    keywordstyle=\color{magenta},
    numberstyle=\tiny\color{codegray},
    stringstyle=\color{codepurple},
    basicstyle=\ttfamily\footnotesize,
    breakatwhitespace=false,         
    breaklines=true,                 
    captionpos=b,                    
    keepspaces=true,                                 
    showspaces=false,                
    showstringspaces=false,
    showtabs=false,                  
    tabsize=2
}

\usepackage{hyperref}
\usepackage{authblk}
\usepackage[round]{natbib}
\usepackage{caption}

\newcommand{\Hyperiax}{\texttt{Hyperiax}}
\newcommand{\JAX}{\texttt{JAX}}

\title{\vspace{-1.5em}\textbf{Hyperiax and Phylogenetic Inference from Shape Data}\vspace{-0.5em}}

\author[1]{Gefan Yang}
\author[ ]{Marcus Teller}
\author[2]{Christy Hipsley}
\author[3,4,5]{Rasmus Nielsen}
\author[1,*]{Stefan Sommer}

\affil[1]{\small Department of Computer Science, University of Copenhagen, Copenhagen 1350, Denmark}
\affil[2]{\small Department of Biology, University of Copenhagen, Copenhagen 2200, Denmark}
\affil[3]{\small GeoGenetics Centre, Globe Institute, University of Copenhagen, Copenhagen 1350, Denmark}
\affil[4]{\small Department of Integrative Biology, University of California, Berkeley, CA 94720, USA}
\affil[5]{\small Department of Statistics, University of California, Berkeley, CA 94720, USA}
\affil[*]{\small Corresponding author}

\date{}

\begin{document}
\twocolumn[
  \begin{@twocolumnfalse}
    \maketitle
    \vspace{-1.5em}
    \begin{abstract}
    \noindent
    \textbf{Summary:} Phylogenetic inference on high-dimensional morphological traits requires algorithms that account for both the nonlinear geometry of the shape data and the phylogenetic tree structure. The Backward Filtering Forward Guiding (BFFG) framework provides smoothing for nonlinear stochastic processes on trees and enables inference of parameters and ancestral states. As practical adoption has been limited by a lack of efficient implementations, we present \Hyperiax, an open-source library for tree traversal algorithms and message passing using \JAX, designed particularly to support operations needed for BFFG. \Hyperiax\ enables efficient execution of operations on trees with large numbers of nodes and, coupled with the BFFG-specific operations, this allows efficient inference in both discrete-time and stochastic differential equation models. Concretely, we demonstrate that \Hyperiax\ enables parameter inference and ancestral reconstruction for butterfly wing shapes represented by landmarks in two dimensions, and analyses of avian beaks from landmarks in three dimensions. Both cases demonstrate application of BFFG on two substantially larger phylogenetic trees with 850 and 696 nodes with higher resolution shape data (118 two-dimensional landmarks and 79 three-dimensional landmarks, specifically) than previously possible.

    \noindent\textbf{Availability and Implementation:} \Hyperiax\ source code and documentation are available at \href{https://github.com/ComputationalEvolutionaryMorphometry/hyperiax}{GitHub} and \href{https://computationalevolutionarymorphometry.github.io/hyperiax/}{GitHub Pages}. Analysis scripts and configuration files are available separately at \href{https://github.com/ComputationalEvolutionaryMorphometry/hyperiax-phylo}{hyperiax-phylo}.

    \noindent\textbf{Contact:} \href{mailto:sommer@di.ku.dk}{sommer@di.ku.dk}

    \vspace{1em}
    \end{abstract}
  \end{@twocolumnfalse}
]

\section{Introduction}
Understanding the evolution of morphological shape requires joint inference of ancestral traits, evolutionary parameters, and their uncertainties on phylogenetic trees. Landmark-based representations capture biologically meaningful geometry but induce strong correlations that challenge classical parsimony or low-dimensional projections \citep{stroustrup2026stochastic}, making it nontrivial to perform statistical inference in the resulting nonlinear stochastic process models. The Backward Filtering Forward Guiding (BFFG) framework \citep{vandermeulen2025backward} overcomes these challenges by combining backward information filters with guided forward simulation. From a broad perspective, the backward passes correspond to the classical Felsenstein tree-pruning algorithm \citep{felsenstein1985phylogenies}, while the forward sampling accounts for the nonlinearity that distinguishes the model from a linear Gaussian model. Practical adoption of BFFG in phylogenetic inference requires efficient numerical schemes and implementations that scale with both tree size and the resolution of the shape data.

\begin{figure}[t]
\centering
\includegraphics[width=0.8\linewidth]{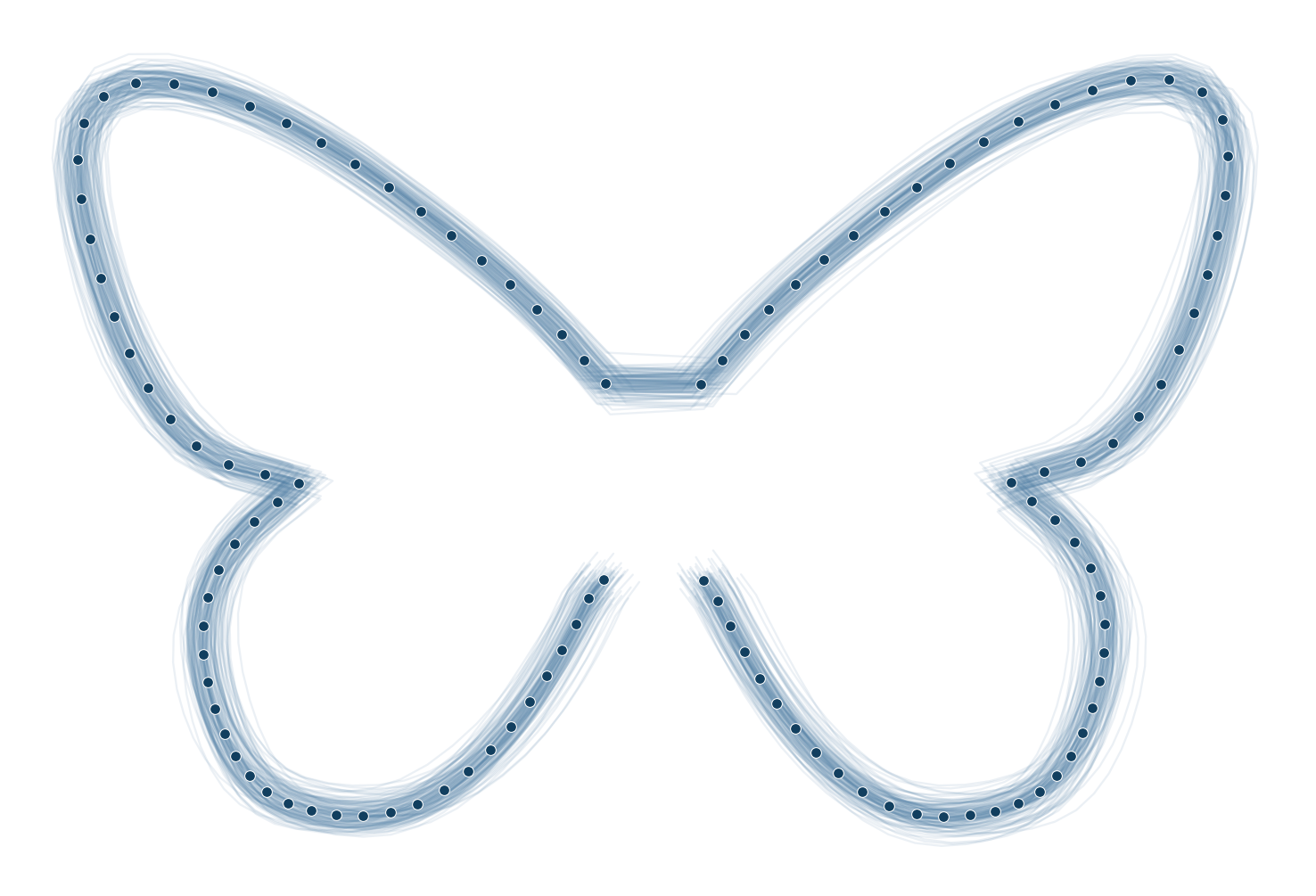}
\caption{Bayesian inference of the phylogenetic root shape for the butterfly dataset described in the Experiments section. The tree has 425 shape observations at the leaves, each represented by 118 landmarks in 2D. Pale blue curves show posterior MCMC samples of the ancestral root shape, while dark points show the posterior mean landmark positions. Inference is performed by repeated \Hyperiax\ tree traversals implementing the BFFG backward-filtering and forward-guiding operations.}
\label{fig:intro}
\end{figure}

The \Hyperiax\ library addresses the above issues. Developed for algorithms requiring tree traversals but specifically optimised for evolutionary shape analysis, the library provides a framework for tree traversal algorithms using Python and \JAX\ \citep{bradbury2018jax} for efficient execution. In this application note, we summarise the architecture of \Hyperiax\ and how it supports the BFFG operations; exemplify the framework with inference from shape data in two and three dimensions; and demonstrate the scalability of the framework by applying it to much larger trees than previously possible. As an example, the posterior distribution of the ancestral root shape of butterfly wings from a phylogenetic tree with 425 observed leaf values is shown in Figure~\ref{fig:intro}. We start with a brief overview of phylogenetic inference from shape data and the BFFG method, before describing \Hyperiax\ and presenting examples.

\section{Phylogenetic inference from morphological data and BFFG}
Shape data can be represented discretely in the form of landmark configurations. For fixed $n>0$, configurations of $n$ distinct landmarks in $\mathbb{R}^d$, $d=2$ or $3$, are represented as points in $\mathbb{R}^{dn}$. This is the case both for Kendall's shape spaces \citep{kendall1984shape} and for diffeomorphism-based landmark models \citep{joshi2000landmark}. Close landmarks will exhibit correlated changes, and, to model this explicitly, in particular with many closely distributed landmarks on the shape, stochastic shape models \citep{stroustrup2026stochastic,sommer2026stochastics} evolve landmarks with a diffusion process whose covariance depends on the current distances between landmarks. The displacement of nearby landmarks is therefore strongly coupled, while distant landmarks can move more independently. Since the diffusion is generated by stochastic deformations of the ambient coordinate system, the evolution is diffeomorphic: non-crossing curves or surfaces remain non-crossing under the process. This gives a parametric evolutionary model for shape, where the amplitude of the covariance kernel controls evolutionary rate and the kernel width controls the spatial scale of morphological integration.

In a phylogenetic context, the stochastic changes over time are governed by the underlying phylogenetic tree, and morphological observations are often at leaves of the tree corresponding to current time. If the transition kernels for subtrees were available in closed form as in the constant-covariance Gaussian case, the likelihood and the conditional process could be computed by a Felsenstein-style backward recursion. However, for general stochastic processes, in particular nonlinear stochastic processes modelling shape evolution, both transition densities and exact backward messages are intractable. BFFG \citep{vandermeulen2025backward} replaces these messages by tractable approximations, typically Gaussian messages obtained from simplified local dynamics. A backward pass propagates the messages \emph{up} along edges from leaves towards the root and \emph{fuses} the information at internal nodes, forming the upward tree traversal\footnote{The naming refers to the computer science convention of the root at the top of the tree and leaves at the bottom of the tree.}. A forward pass then simulates a guided process from the root towards the leaves, obtained by adding a drift term depending on the approximate backward information. This downward traversal is defined by \emph{down} operations at each edge. The resulting paths are weighted by a Radon-Nikodym correction, so they can be used in importance sampling or Metropolis-Hastings updates for ancestral shapes and model parameters. Thus, the fundamental operations of the method are a repeated sequence of tree traversals governed by \emph{up} and \emph{down} operations and local message \emph{fusion}, which exactly corresponds to the operations implemented in \Hyperiax. BFFG supports both discrete- and continuous-time transitions.

\section{Hyperiax}
\Hyperiax\ treats tree algorithms as \JAX\ transformations: users compose them from local \emph{up}, \emph{down}, and fusion operations, while the library handles traversal order, ragged child sets, batching, just-in-time (JIT) compilation, automatic differentiation, and accelerator execution. For phylogenetic shape analysis, this makes the repeated BFFG backward-filtering and forward-guiding passes executable as compiled tree sweeps rather than as hand-written Python loops.

\subsection{Library architecture}
\Hyperiax\ is organized around a small set of abstractions. A \texttt{Topology} stores the rooted tree itself, including parent-child relationships, traversal order, and level structure; a \texttt{Tree} stores typed arrays attached to the nodes of that topology. This separation mirrors the distinction between a phylogenetic tree and the quantities propagated over it, such as shape states, likelihoods, and correction weights. Computations are written as local message updates and lifted to whole-tree traversals through the \texttt{@up} and \texttt{@down} decorators. An up-sweep corresponds to a postorder pass from the tips to the root, while a down-sweep propagates information from ancestors to descendants. Each sweep declares which fields it reads and writes. The following schematic example shows how these abstractions appear in a typical setup, starting from a Newick tree and defining one upward and one downward pass.

\begin{lstlisting}[language=Python]
import hyperiax as hx

# Read a phylogeny from Newick and allocate per-node data fields.
tree = hx.from_newick(
    "tree.nwk", # Newick file
    schema={
        "likelihood": (), # scalar
        "state": (state_dim,), # state vector
    },
)

topo = tree.topology
# Initialize tree with JAX "out-of-place" updates.
tree = tree.at[topo.is_leaf].set(
    likelihood=leaf_likelihoods
)
tree = tree.at[topo.is_root].set(
    state=root_state[None, :]
)

@hx.up(
    reads_children=("likelihood",), 
    writes=("likelihood",)
)
def pruning(node, children, params):
    # Upward sweep: 
    # combine child likelihoods at an internal node by production.
    return {"likelihood": children.likelihood.prod(0)}

@hx.down(
    reads_parent=("state",), 
    writes=("state",)
)
def inherit(node, parent, params):
    # Downward sweep: 
    # propagate ancestral state directly to descendants.
    return {"state": parent.state}

# Execute sweeps, return new Trees
pruned_tree = pruning(tree) # tips -> root
inherited_tree = inherit(pruned_tree) # root -> tips
\end{lstlisting}

Internally, \Hyperiax\ uses \JAX\ immutable data structures, so sweeps are \texttt{Tree} $\to$ \texttt{Tree} transformations that can be compiled, differentiated, and repeated inside \JAX\ loops. \Hyperiax\  handles both balanced and ragged phylogenies. It supports vectorized reductions over children and still avoids dense padding for unbalanced trees. The \texttt{core} module of the library provides topology construction, schemas, tree containers, views, and sweep execution; \texttt{utils} contains \JAX\ numerical routines; and \texttt{prebuilt} contains components including weighted phylogenetic means and the BFFG operations for both discrete and continuous transition models. The design aims to let users write algorithms close to their mathematical formulation, while the library handles traversal, batching, JIT compilation, and execution underneath.

\subsection{BFFG-specific components}
\Hyperiax\ provides BFFG-specific components in the \texttt{prebuilt} module. Backward filtering is an \emph{up} sweep from observed tips to the root; child messages are fused by a local reduction at each internal node; forward guiding is a \emph{down} sweep from the root back to the tips. To define a specific model, one defines discrete transition means and covariances or continuous-time drift and diffusion terms, together with tractable auxiliary approximations.

The BFFG state is stored directly as fields on a \texttt{Tree}: shape states (single shape or time-parametrized paths of shapes), Gaussian message parameters, random driving noise, normalizing terms, and per-edge correction weights. Inference in the form of sampling parameters from the posterior distribution can be done in an MCMC loop iterating up and down sweeps where the Metropolis--Hastings accept/reject decisions are based on the computed normalizing terms and correction weights.

Both discrete- and continuous-time models use the same structure but with different edge updates. For discrete Gaussian transitions, the backward sweep propagates Gaussian messages in canonical form under a local auxiliary model, and the forward sweep samples each node conditional on its parent and the downstream messages. In the case of linear-Gaussian transitions with constant covariance, the backwards messages are exact and the importance correction is zero, which gives the closed-form recursion. For continuous-time shape models, the edge updates use an auxiliary SDE. For the shape examples specifically, the diffusion matrix for the auxiliary process is computed at an approximation of the shape state at the inner nodes.

\subsection{Other applications}
\Hyperiax\ is a general framework for algorithms requiring tree traversals. In the phylogenetic context, besides BFFG, phylogenetic mean estimation is supported through upward recursion, for example for shape data as pursued by \citet{severinsen2026diffeomorphic}.

\section{Examples}
\label{sec:examples}
We exemplify parameter inference from shape data with \Hyperiax\ and the BFFG operations on butterfly and beak shape data. Table~\ref{tab:mcmc-summary} summarizes the dataset sizes, posterior parameter estimates, and convergence diagnostics for both examples.

\begin{table*}[h]
\centering
\small
\setlength{\tabcolsep}{4pt}
\caption[Dataset info and MCMC summaries for the two shape examples.]{Dataset scale and MCMC summaries for the two shape examples. Posterior summaries are reported as posterior mean [95\% credible interval] from four independent chains of 5000 iterations, with the first 500 iterations of each chain discarded as burn-in. Here $k_\alpha$ controls diffusion amplitude (diffusivity), $k_\sigma$ the spatial correlation scale of landmark motion, and $\sigma^2_{\mathrm{obs}}$ observation variance. The final column reports the maximum Gelman--Rubin $\hat R$ over these three parameters. See \cite{stroustrup2026stochastic} for more information about the shape process parameters.}
\begin{tabular}{@{}llrrllll@{}}
\toprule
Dataset & Shape representation & Tips & Nodes & $k_\alpha$ & $k_\sigma$ & $\sigma^2_{\mathrm{obs}}$ ($10^{-3}$) & $\hat R_{\max}$ \\
\midrule
Butterfly wings & 118 2D landmarks & 425 & 850 & 0.134 [0.106, 0.167] & 0.212 [0.178, 0.242] & 1.25 [0.99, 1.56] & 1.002 \\
Avian beaks & 79 3D landmarks & 348 & 696 & 0.113 [0.080, 0.157] & 0.152 [0.121, 0.194] & 0.408 [0.287, 0.580] & 1.011 \\
\bottomrule
\end{tabular}
\label{tab:mcmc-summary}
\end{table*}

\subsection{Butterfly wing landmarks (2D)}
\begin{figure}[t]
\centering
\begin{minipage}[t]{0.48\linewidth}
  \raggedright\textbf{(a)}\par\vspace{0.2em}
  \centering
  \vspace{.5cm}
  \includegraphics[width=\linewidth]{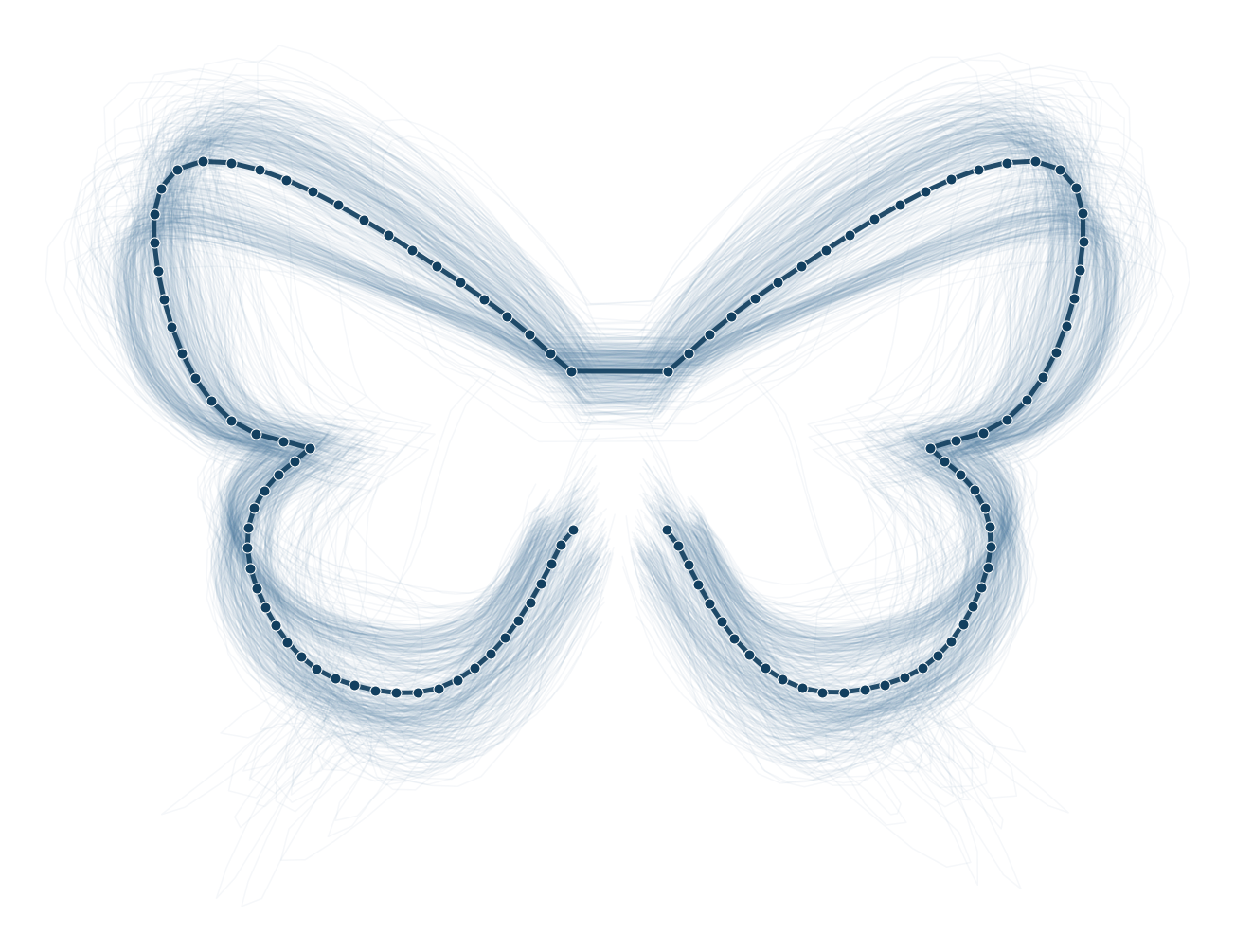}
\end{minipage}
\hfill
\begin{minipage}[t]{0.48\linewidth}
  \raggedright\textbf{(b)}\par\vspace{0.2em}
  \centering
  \includegraphics[width=\linewidth]{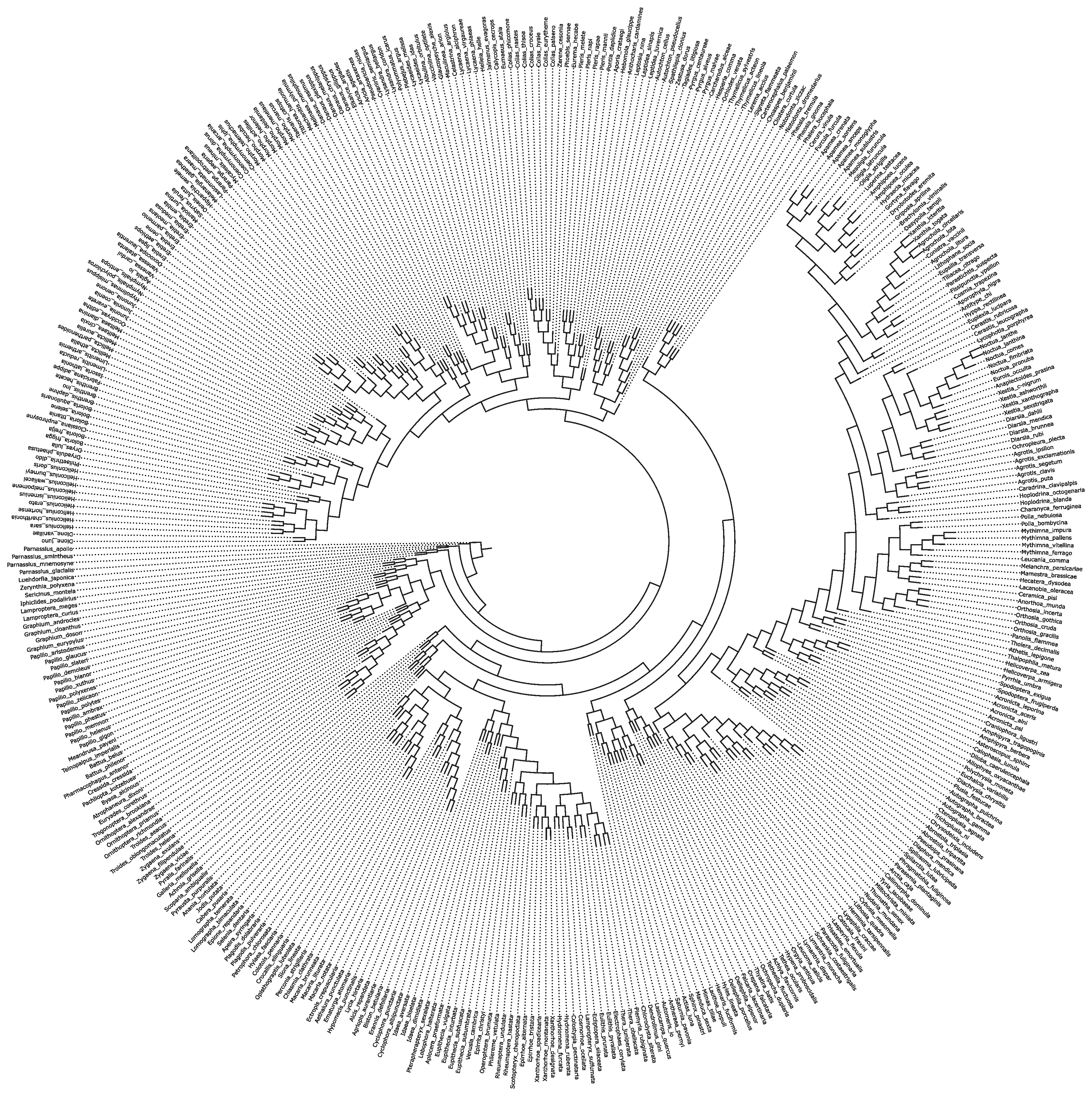}
\end{minipage}

\vspace{0.75em}
\begin{minipage}[t]{0.48\linewidth}
  \raggedright\textbf{(c)}\par\vspace{0.2em}
  \centering
  \includegraphics[width=\linewidth]{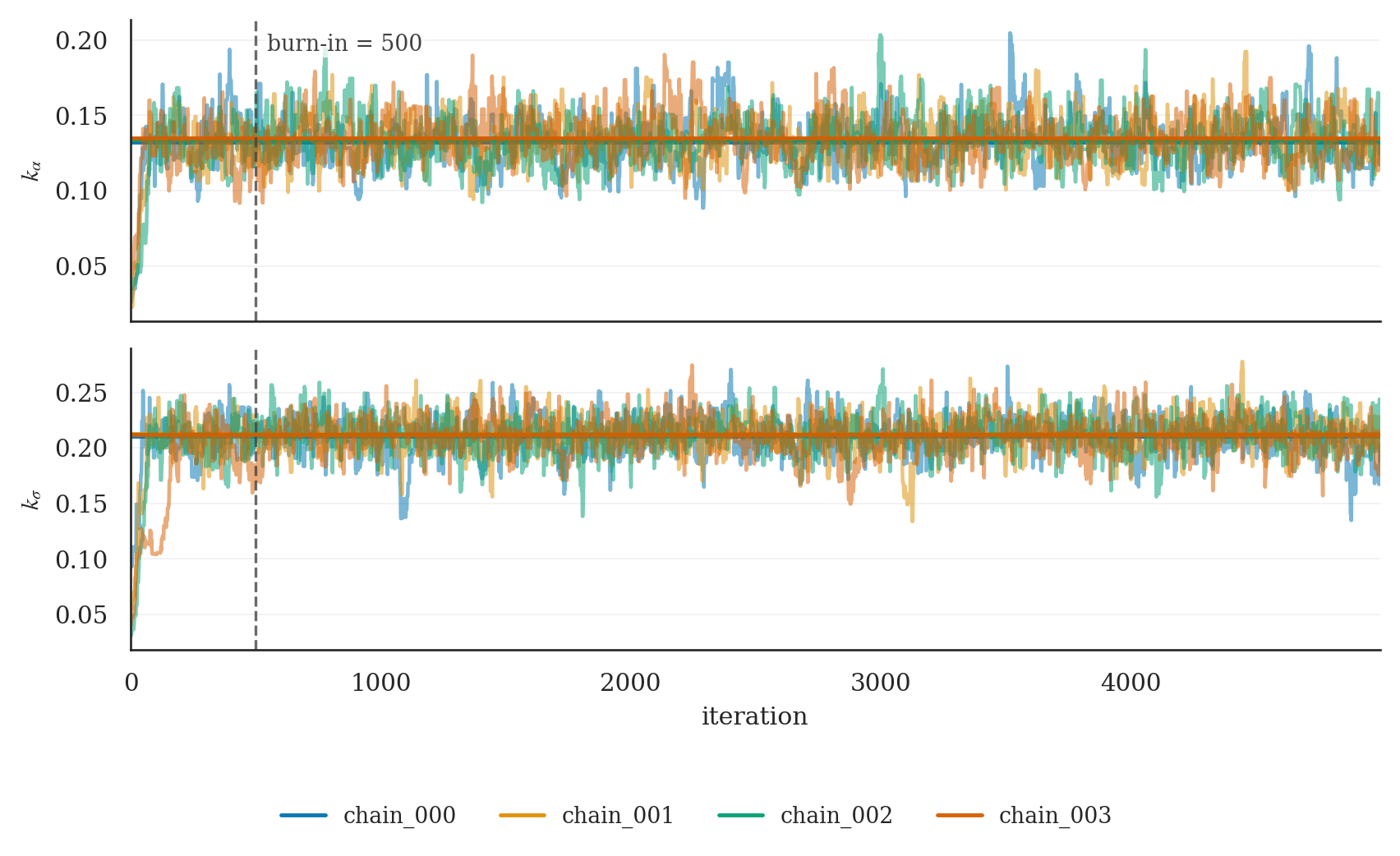}
\end{minipage}
\hfill
\begin{minipage}[t]{0.48\linewidth}
  \raggedright\textbf{(d)}\par\vspace{0.2em}
  \centering
  \includegraphics[width=\linewidth]{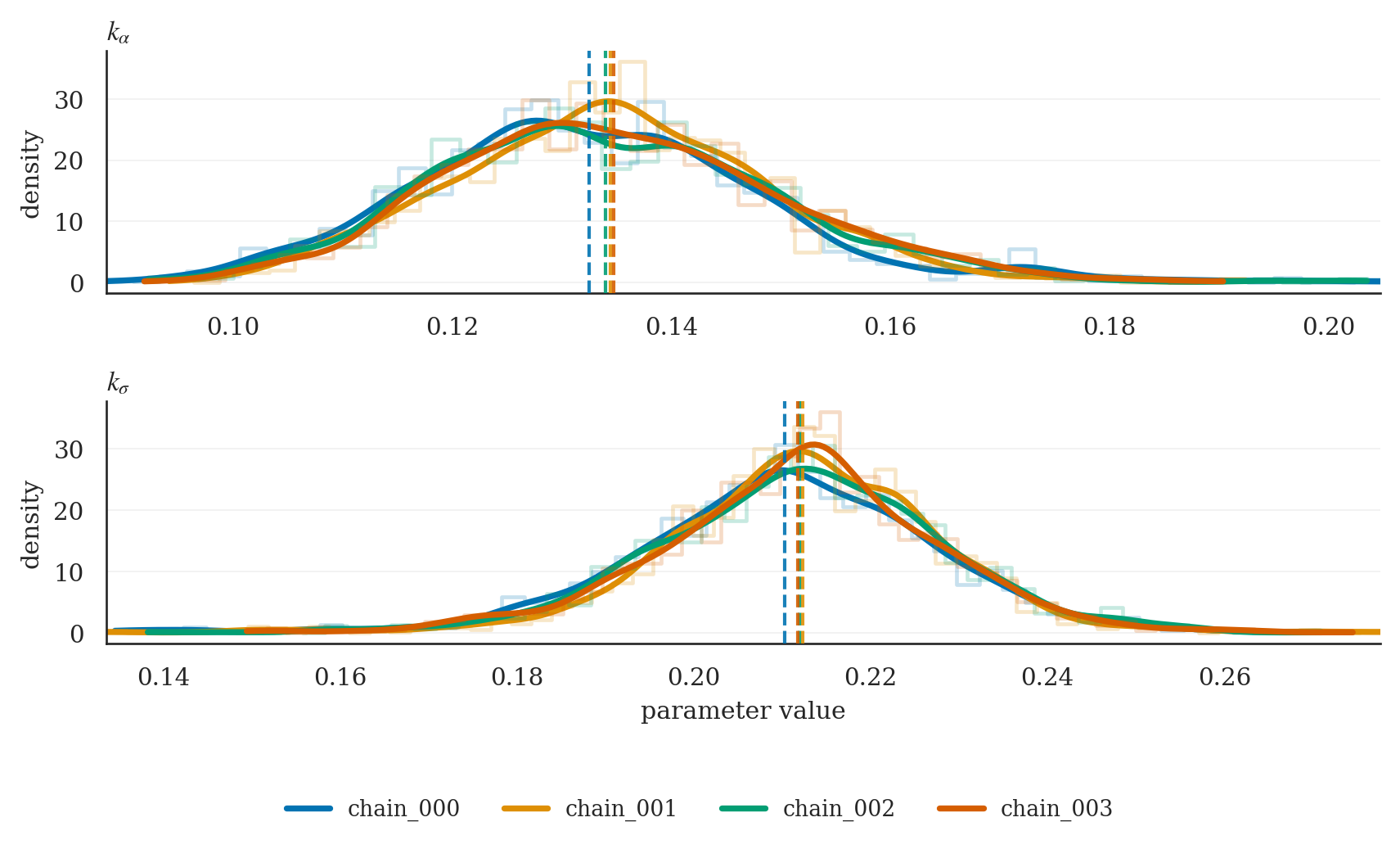}
\end{minipage}
\caption{Inference of parameters and ancestral states for butterfly shapes. (a) Observed leaf-shape ensemble across all 425 leaves; pale curves show individual leaf shapes and dark landmarks summarize the mean observed configuration. (b) Phylogenetic tree. (c) MCMC traces for the parameters of the shape process - diffusivity $k_\alpha$ and spatial correlation $k_\sigma$. (d) Marginal posterior densities for the same parameters. Dashed vertical lines and labels mark posterior means.}
\label{fig:butterflies}
\end{figure}

We use the stochastic model of \citet{stroustrup2026stochastic} to model the temporal stochastic dynamics of butterfly wings represented by 118 landmarks. Current time observations correspond to the leaf shapes. We use \Hyperiax\ to load the topology of the phylogenetic tree, perform the tree passes, and run the MCMC sampler. Input data, tree, and results are shown in Figure~\ref{fig:butterflies}, including MCMC traces and posterior distributions of the parameters of the shape process that govern the diffusivity in the model, the spatial correlation, and the observation noise. Figure~\ref{fig:intro} additionally shows posterior samples from the root of the tree.

The phylogenetic tree contains 425 leaves and 850 nodes in total, thereby extending the typical previous scale of application of the BFFG method, on the order of 30 leaves and 30 internal nodes \citep{stroustrup2026stochastic}. The dataset is further described in \cite{severinsen2026stochastic}.

\subsection{Avian beak landmarks (3D)}
\begin{figure}[t]
\centering
\begin{minipage}[t]{0.48\linewidth}
  \raggedright\textbf{(a)}\par\vspace{0.2em}
  \centering
  \includegraphics[width=\linewidth,clip=True,trim=50 50 50 50]{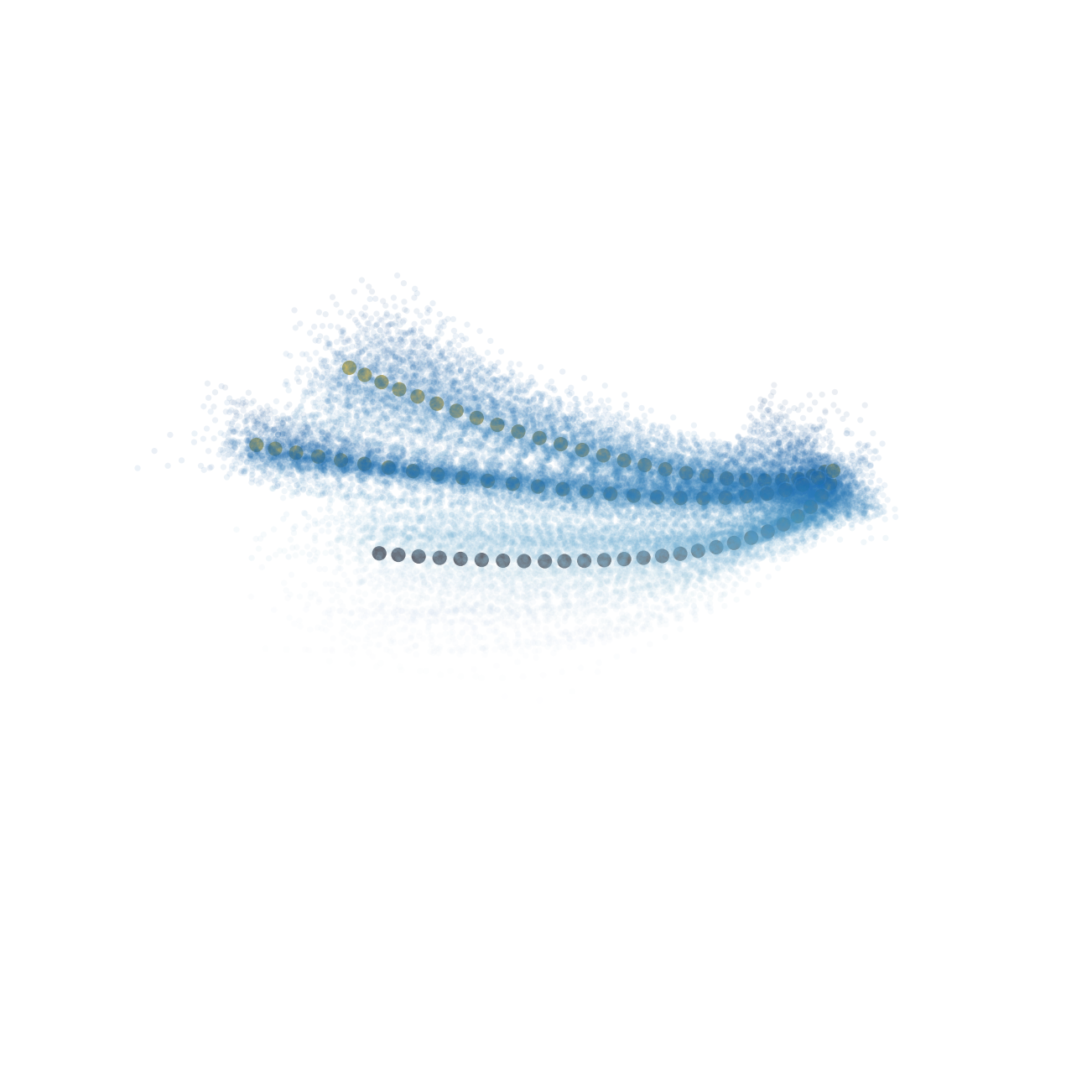}
\end{minipage}
\hfill
\begin{minipage}[t]{0.48\linewidth}
  \raggedright\textbf{(b)}\par\vspace{0.2em}
  \centering
  \includegraphics[width=\linewidth]{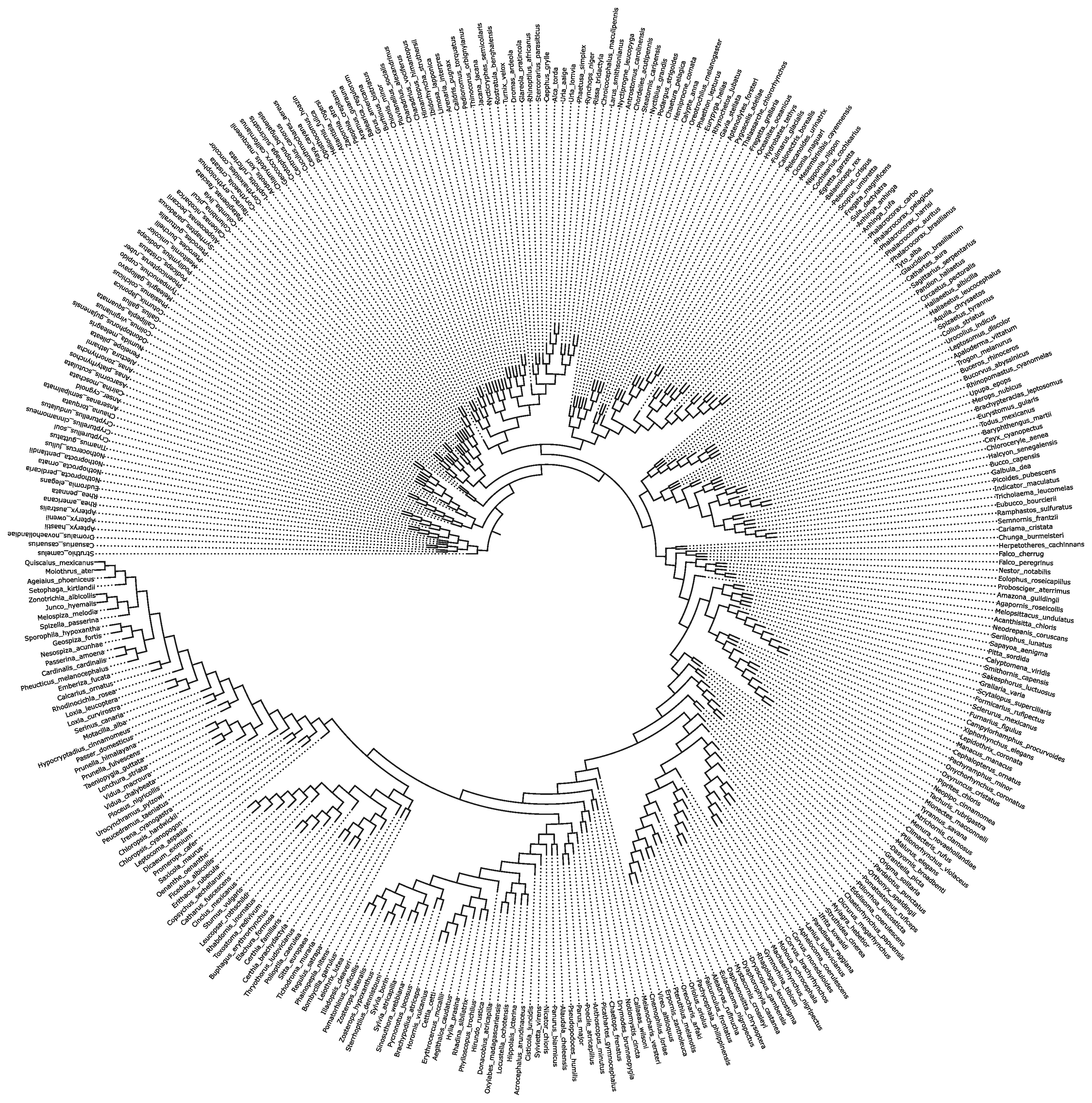}
\end{minipage}

\vspace{0.75em}
\begin{minipage}[t]{0.48\linewidth}
  \raggedright\textbf{(c)}\par\vspace{0.2em}
  \centering
  \includegraphics[width=\linewidth]{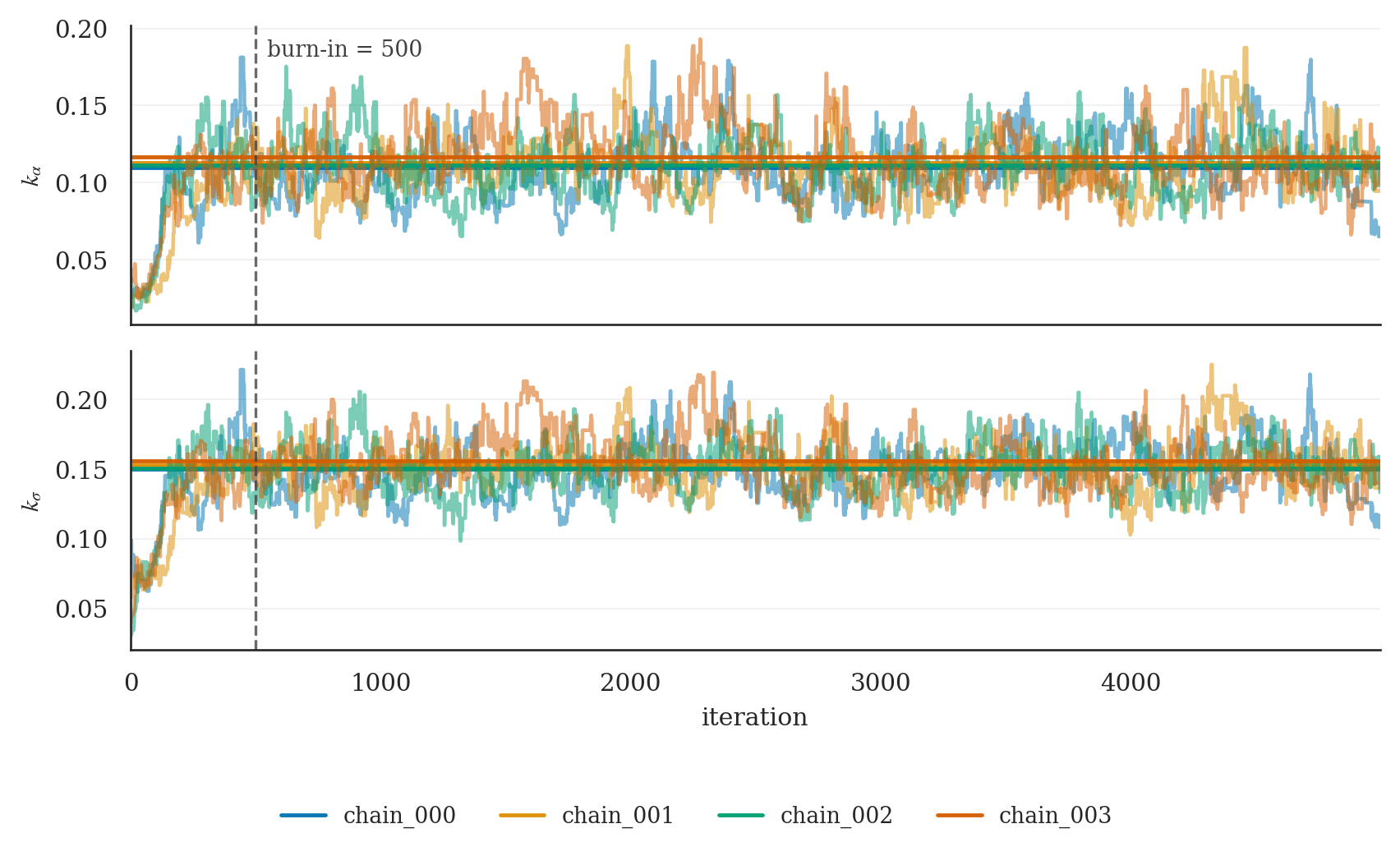}
\end{minipage}
\hfill
\begin{minipage}[t]{0.48\linewidth}
  \raggedright\textbf{(d)}\par\vspace{0.2em}
  \centering
  \includegraphics[width=\linewidth]{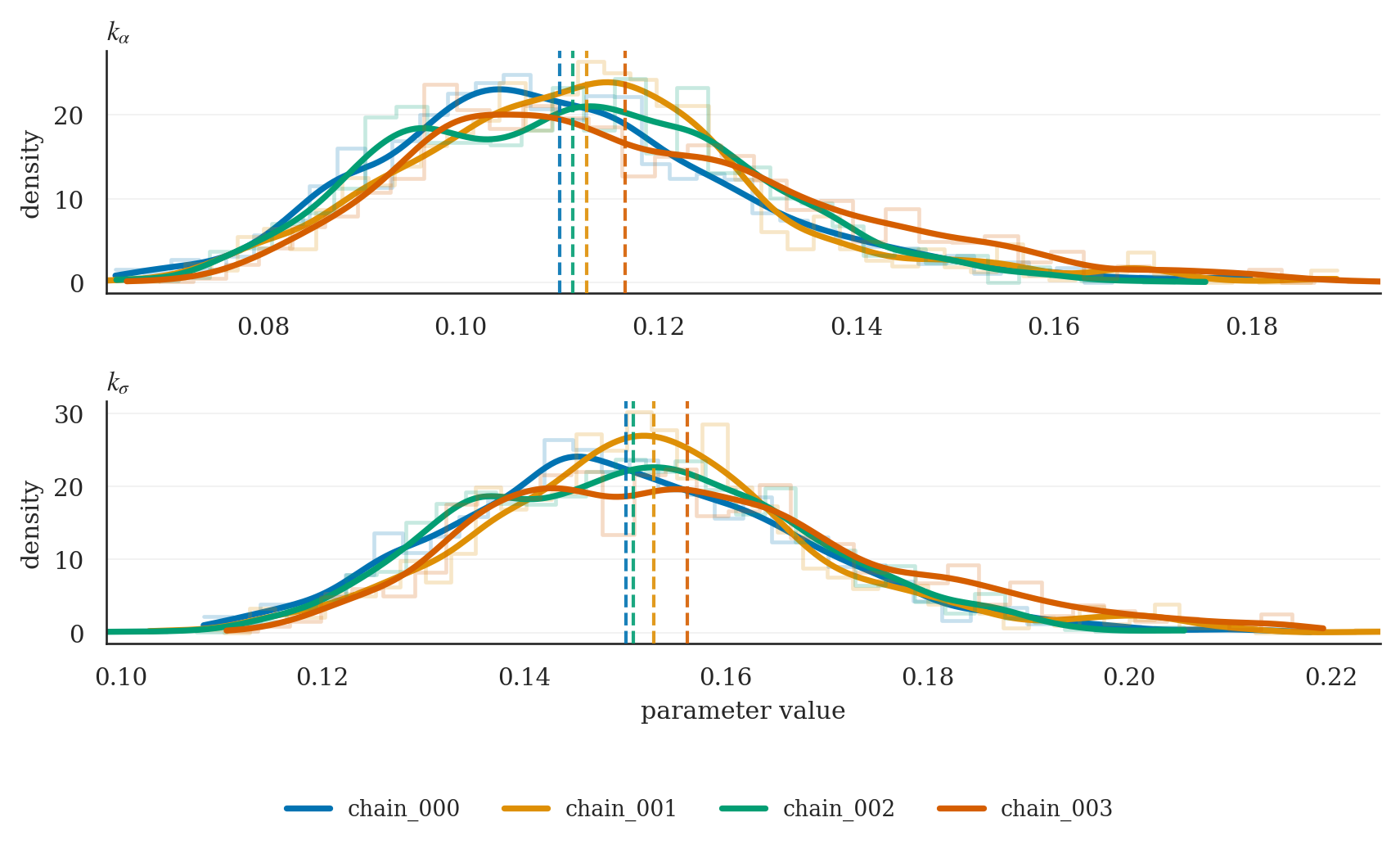}
\end{minipage}
\caption{Avian beak example. (a) Observed leaf shapes across the 348 leaves, represented by 79 three-dimensional landmarks; pale points show landmarks of individual leaf shapes and dark points summarize the mean observed landmarks. (b) Phylogenetic tree. (c) MCMC traces for the shape process parameters - diffusivity $k_\alpha$ and spatial correlation $k_\sigma$. (d) Marginal posterior densities for the parameters. Dashed vertical lines and labels mark posterior means.}
\vspace{-10pt}
\label{fig:beaks}
\end{figure}

We repeat the butterfly experiment with three-dimensional landmarks sampled on avian beaks. The results are displayed in Figure~\ref{fig:beaks}, again showing estimates of the parameters of the model. The landmarks are from the \url{MarkMyBird.org} citizen science project first described in \citet{cooney2017mega} and the phylogenetic tree is described in \citet{stiller2024complexity}. This experiment demonstrates the application of BFFG for 3D data and large trees.

\section{Discussion}
\Hyperiax\ provides a direct interface for applying the BFFG operations to phylogenetic shape inference problems, and does so efficiently and scalably using \JAX\ for parallelization, just-in-time compilation and automatic differentiation. We demonstrate this by performing parameter inference in phylogenetic models for 2D and 3D shape data. Using \Hyperiax, we are able to scale to higher shape resolution (number of landmarks) and larger trees than previously possible.

\section*{Availability and reproducibility}
The source code of \Hyperiax\ is available at \href{https://github.com/ComputationalEvolutionaryMorphometry/hyperiax}{GitHub}, with documentation hosted at \href{https://computationalevolutionarymorphometry.github.io/hyperiax/}{GitHub Pages}. The analyses in this application note are available in a separate reproduction repository, \href{https://github.com/ComputationalEvolutionaryMorphometry/hyperiax-phylo}{\texttt{hyperiax-phylo}}, which contains the scripts for performing the described experiments, configuration files, data-processing workflow, and figure-generation code. The scripts import \Hyperiax\ as a dependency.

\section*{Funding}
This work was supported by the Villum Foundation Grant 40582, the Novo Nordisk Foundation grants NNF18OC0052000, NNF24OC0093490 and NNF24OC0089608.

\section*{Conflict of interest}
The authors declare no conflicts of interests.

\bibliographystyle{plainnat}
\bibliography{references}

\end{document}